\documentclass[twocolumn, aps, prl, amsmath, amssymb, reprint, superscriptaddress]{revtex4-1}
\usepackage{graphicx}
\usepackage{dcolumn}
\usepackage{bm}
\usepackage[utf8]{inputenc}
\usepackage[T1]{fontenc}
\usepackage{mathptmx}
\usepackage[version=4]{mhchem}
\usepackage[dvipsnames]{xcolor}

\newcommand{\eg}{e.g., }
\newcommand{\ie}{i.e., }

\newcommand{\Eq}[1]{Eq.~(\ref{#1})}

\begin{document}

\title{Autoresonant Removal of Fusion Products in Mirror Machines}

\author{Eli Gudinetsky}
\affiliation{Racah Institute of Physics, The Hebrew University of Jerusalem, Jerusalem, 91904 Israel}

\author{Tal Miller}
\affiliation{Racah Institute of Physics, The Hebrew University of Jerusalem, Jerusalem, 91904 Israel}

\author{Ilan Be'ery}
\affiliation{nT-Tao, 5 Ha-Nagar st., Ramat Hasharon, 4526005 Israel}

\author{Ido Barth}
\email{ido.barth@mail.huji.ac.il}
\affiliation{Racah Institute of Physics, The Hebrew University of Jerusalem, Jerusalem, 91904 Israel}

\date{\today}


\begin{abstract}
Magnetic confinement fusion reactors produce fusion byproduct particles that must be removed for efficient operation.
It is suggested to use autoresonance (a continuous phase-locking between anharmonic motion and a chirped drive) to remove the fusion products from a magnetic mirror, the simplest magnetic confinement configuration. 
An analogy to the driven pendulum is established via the guiding center approximation.
The full 3D dynamics is simulated for $\alpha$ particles (the products of DT fusion) in agreement with the approximated 1D model.
Monte Carlo simulations sampling the phase space of initial conditions are used to quantify the method's efficiency.
The DT fuel particles are out of the bandwidth of the chirped drive and, therefore, stay in the mirror for ongoing fusion.
The method is also applicable for advanced, aneutronic reactors, such as p-$^{11}$B.
\end{abstract}

\maketitle


The most straightforward magnetic confinement configuration is the magnetic mirror and thus serves as the basic building block for various linear fusion machines that are advantageous not only for their engineering simplicity but also for their high-$\beta$ steady-state operation \cite{post1987magnetic}.
The fusion products, \eg $\alpha$ particles in deuterium-tritium (DT) fusion, are trapped by the same mirroring magnetic field that traps the fuel particles, taking up the place of the valuable fuel particles.
The removal of fusion products poses a long-standing problem \cite{khvesyuk1995ash,fisch2006alpha,Fetterman_Fisch.PRL.2008,reiter1991helium,khvesyuk1995ash,white2021alpha,bierwage2022energy,Fisch.PRL.1992,Zhmoginov_Fisch_pop_2008,fisch1994alpha,mynick1994frequency}. 
One of the leading methods is the $\alpha$-channeling, which employs rf-waves to induce directed diffusion of the $\alpha$ particles while cooling down and transferring their energy through the plasma waves to the fuel particles \cite{Fisch.PRL.1992,Zhmoginov_Fisch_pop_2008}.
Retaining the energy of the $3.5$~MeV $\alpha$ particles in the plasma is beneficial for a self-sustaining burning process \cite{Lawson1957,Wurzel2022}. 
However, further heating beyond $15$~keV is counterproductive for DT fusion due to decreased reactivity and increased bremsstrahlung radiation losses \cite{Wurzel2022}.
Thus, the removal of the high-energy $\alpha$ particles from the fusion cell with (part of) their energy could be advantageous for a continuous temperature and burn control, particularly if the energy of the $\alpha$ beam is extracted by direct energy conversion outside the mirror \cite{Moir1973,barr1982experimental}. 

Autoresonance (AR), \ie a continuous phase locking in a nonlinear oscillatory system with slowly varying parameters, has been thoroughly studied theoretically 
\cite{meerson1990strongT,liu1995nonlinearT,Friedland1998T,fajans2001autoresonantT,Friedland_2008,barth2009autoresonantT,barth2014quantumT,Armon2016T} and experimentally \cite{fajans1999autoresonantE,naaman2008phaseE,barak2009autoresonantE,murch2011quantumE,andresen2011autoresonantE,shalibo2012quantumE} in various systems, where the simplest example is the chirped-driven pendulum \cite{Lazar_proc_2005}.
Phase-locking is established when the driving frequency passes through the linear resonance and is preserved as the nonlinear frequency shift compensates for the chirped driving frequency.
AR provides control of the oscillating degree of freedom without the need for feedback and, thus, can be used to release charged particles from a trapping potential, \eg anti-protons from a Penning trap for anti-hydrogen production \cite{andresen2011autoresonantE,AntiHydrogen_nature_2010}.
The capture into resonance of the chirped-driven pendulum has two different limits.
The first is the probabilistic capture, where the oscillating particle begins at a large amplitude state \ie with a sufficiently large nonlinear (anharmonic) frequency shift. 
In this regime, only a small fraction of the initial oscillation phases is captured into resonance, while all other particles experience a transient resonant kick and continue in a nonresonant motion \cite{Neishtadt_PRE_2005,Armon2016T}.
The frequency sweeping technique for toroidal systems \cite{mynick1994frequency} operates in such a regime.
The second limit is the automatic capture, where almost all particles near equilibrium are captured into AR regardless of their initial phase. 
The capture in this case occurs only if the driving amplitude, $\varepsilon$, exceeds a critical value, $\varepsilon_\text{cr}$ that scales as $\alpha^{3/4}$, where $\alpha$ is the chirp rate \cite{Friedland_2008,fajans1999autoresonantE,barth2009autoresonantT}.
Notably, the captured particles may reach large amplitudes as the chirping of the driving frequency continues.

In this letter, we study AR in a magnetic mirror for the first time and propose its application for the quick removal of a large fraction of fusion products.
The idea is to autoresonantly increase the longitudinal energy of the $\alpha$ particles until they escape through the loss cone.
The driving force is realized via a weak oscillating axial magnetic field with a slowly down-chirped frequency that passes through the linear bouncing frequency.
As a result, most of the $\alpha$ particles near the machine's center are phase-locked with the drive and gain axial velocity until they escape through the loss cone. 
Such an AR cycle can be successively repeated for continuous product removal in burning plasmas.
Crucially, the $\alpha$ particles are removed without significantly affecting the fuel particles.


Consider a charged particle with mass $m$ and charge $q$ trapped in a magnetic mirror of length $2l$ and driven by a slowly chirped oscillatory field.
We approximate the longitudinal magnetic field near the mirror axis in cylindrical coordinates, $(r,\theta,z)$, as
\begin{eqnarray}\label{eq:B_z}
    B_z=B_\text{min}+B_0\left(1-\cos\ \frac{\pi z}{l}\right) + \varepsilon B_0\frac{\pi z}{l}\cos{\phi_\text{d}}
\end{eqnarray}
where $B_0=\left(B_\text{max}-B_\text{min}\right)/2$ and the maximum (minimum) static magnetic field is $B_\text{max}$ $(B_\text{min})$.
The last term in \Eq{eq:B_z} is the time-dependent driving field, where $\varepsilon\ll1$ is the dimensionless amplitude and $\phi_\text{d}=\int\omega_\text{d} dt$ is the phase of the chirped frequency, $\omega_\text{d}=\omega_0-\alpha\omega_0^2\, t$. 
The driving frequency, $\omega_\text{d}$, is chosen such that around $t=0$ it will resonate with the longitudinal ($z-$direction) bouncing frequency, $\omega_B=\sqrt{\mu B_0 \pi^2 /m\,l^2}$ of a typical trapped particle (see Eq.~\ref{eq: pendulum} below).
The magnetic moment, $\mu=0.5mv_{\perp}^2/B_z$ is an adiabatic invariant because $\omega_B,\omega_\text{d}\ll\omega_c$, where $\omega_c$ is the cyclotron frequency. 
The driving field can be realized by oscillating the currents in the mirror coils.
Because of the symmetry of the mirror, the azimuthal magnetic field, $B_\theta$ is zero, while the radial component is determined by
\begin{eqnarray}\label{eq:B_r}
    B_r=- \frac{ \pi r B_0}{2l}\left(\sin\ \frac{\pi z}{l}+\varepsilon\cos{\phi_\text{d}}\right)
\end{eqnarray}
due to the magnetic Gauss' law, $\nabla\cdot \textbf{B}=0$.
The time-varying components of the magnetic field, \ie the driving field, induce an azimuthal electric field,
\begin{eqnarray}\label{eq:E_d}
    \textbf{E}=-\frac{r}{2}\,\frac{\partial B_z}{\partial t}\,\hat{\theta}
    = \frac{\varepsilon \, \pi \,r \, z \, B_0 \, \omega_\text{d}}{2l}  \sin{\phi_\text{d}}\,\hat{\theta}.
\end{eqnarray}


Most notably, the full three-dimensional (3D) dynamics can be reduced into 1D using the guiding center approximation provided $mcv_\perp/qB_0 l \ll 1$.
The approximated longitudinal (along the magnetic field) equation of motion then reads \cite{Northrop_book}

\begin{equation} \label{eq: guiding center}
    m\frac{d v_\parallel}{d t}\approx
    q E_\parallel 
    - \mu\frac{\partial B}{\partial s}
    + m\, \textbf{u}_E\cdot \frac{d \hat{e}_1}{dt}
\end{equation}
where we neglected gravitation.
$E_\parallel$ is the longitudinal components of the electric field evaluated at the guiding center trajectory. 
The derivative $\partial/\partial s$ is along the field line, $\textbf{u}_E$ is the $E\times B$ drift velocity, and $\hat{e}_1$ is a unit vector along the field line.
For a guiding center trajectory on the mirror axis, we can thus substitute $\partial/\partial s = \partial/\partial z$ and $\hat{e}_1=\hat{z}$.
Notably, for the fields of Eqs.~(\ref{eq:B_z}-\ref{eq:E_d}), the only nonvanishing term is $\mu \partial B/ \partial s$ because \Eq{eq:E_d} yields $E_\parallel=0$ and thus also $\textbf{u}_E(r=0)=0$.
Consequently, the guiding center equation for the longitudinal motion of particles near the mirror axis reduces to the well-studied, 1D chirped-driven pendulum  \cite{Lazar_proc_2005,barth2009autoresonantT,Friedland_2008}
\begin{equation} \label{eq: pendulum}
    \frac{d v_\parallel}{d t}=-\frac{\mu B_0\pi}{m l} \left(\sin\frac{\pi z}{l}+\varepsilon \cos \varphi_\text{d} \right).
\end{equation}
Therefore, we expect to observe AR dynamics in magnetic mirrors for suitable parameters.

\begin{figure}[tb]
    \includegraphics[clip, trim=0.6cm 0.1cm 1.0cm 0.6cm,
    width=1\linewidth]{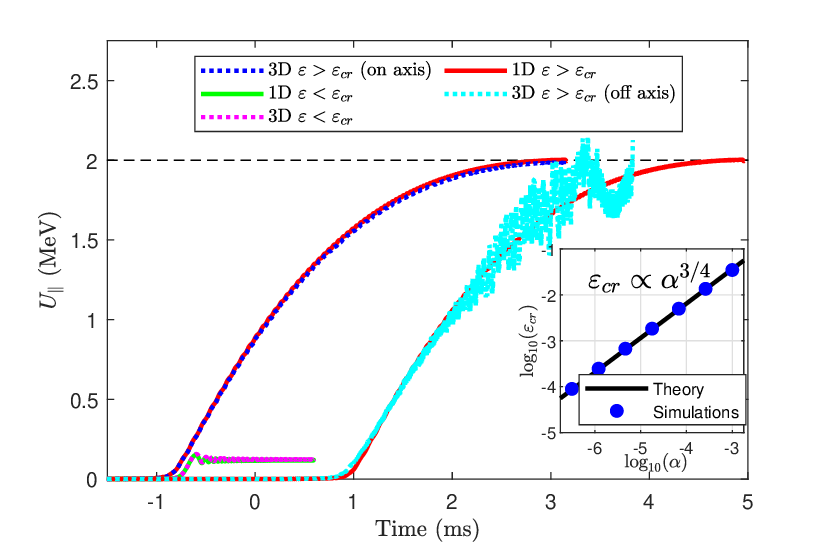}
    \caption{The dynamics of the axial energy (horizontally shifted for clarity), as evaluated at the mirror mid-plane, under the influence of the chirped drive above, $\varepsilon>\varepsilon_\text{cr}$, (red, blue, and cyan) and below, $\varepsilon<\varepsilon_\text{cr}$, (green and magenta) the threshold, for the 1D guiding center approximated model (solid lines) and the 3D simulations (dotted lines). 
    Presented are trajectories with an initial guiding center on the mirror axis (blue and magenta) and at $r_{gc}=0.5$~m off the mirror axis (cyan).
    The dashed black line indicates the escape axial energy for an on-axis particle.
    Inset: The threshold, $\varepsilon_\text{cr}$, versus the chirp rate, $\alpha$, in log-log scales, as found in 3D simulations (dots) in comparison to the theoretical scaling (solid line).
    }
    \label{fig: capture_scatter_power_law}
\end{figure}


To test this prediction, we solve the 3D equations of motion for a nonrelativistic $\alpha$ particle with an initial energy of $1$~MeV under the influence of the magnetic and electric fields in Eqs.~(\ref{eq:B_z}-\ref{eq:E_d}).
Our solver for the exact Lorentz force is fully 3D, utilizing a volume-preserving scheme \cite{he2015volume}.
Targeting this initial energy regime complies with the typical slowed-down mean energy of $\sim1$~MeV \cite{heidbrink1994behaviour, zweben2000alpha} in burning plasmas.
Besides, it is generally advantageous over considering $3.5$~MeV particles because they deposit most of their energy in the plasma before being removed from the mirror and because direct energy conversion methods (outside the mirror) are more efficient for lower energies. 
We consider $l=5$~m, $B_\text{min}=1$~T, and $B_\text{max}=3~$T, which satisfies the guiding center approximation since $mcv_\perp/qB_0 l\approx 5\times10^{-3}\ll1$.
The driving parameters were $\omega_0=2\pi\times 0.8$~MHz, $\alpha=2\times10^{-5}$, and $\varepsilon=8\times10^{-3}$, corresponding to about $25$~mT maximal driving magnetic field at the mirror's throat. 
First, we illustrate the AR dynamics in mirror machines by considering a few examples with near-equilibrium initial conditions. 
Then, we generalize the simulations to the whole phase space to study the efficiency of the AR removal scheme.

In Fig.~\ref{fig: capture_scatter_power_law}, we present the dynamics of two $\alpha$ particles beginning at the mid-plane of the mirror, with zero longitudinal velocity (so $\omega_B\approx 2\pi\times 0.5$~MHz), one with gyro-center located on mirror axis and one off-axis.
The numerical results exhibit a typical AR solution for both on- and off-axis particles (blue and cyan dotted lines, respectively), including the capture into resonance when passing through the linear resonance and the slow modulation of the energy in the nonlinear regime \cite{Lazar_proc_2005}.
The particle escapes through the loss cone after about 3~ms when its parallel energy reaches $2$~MeV.
We compare the full 3D solution with the 1D approximated dynamics of Eq.~(\ref{eq: pendulum}) (solid red lines) for the same parameters and found an excellent agreement at all times for the on-axis case and until the particle is excited to higher energies for the off-axis particle.
Importantly, the magnetic moment is adiabatically conserved. 
Therefore, although the axial energy, $U_\parallel$, increases, the transverse energy (and the radial position) remains bounded (not shown in the figure).
The deviation from the 1D trajectory at high amplitudes is probably because of nonlinear coupling with other degrees of freedom.

In the figure, we also illustrate the dynamics below the AR threshold, where $U_\parallel$ is modestly excited when passing through the linear resonance followed by energy saturation after $\sim0.1$~ms.
The parameters, in this case, were the same as before except the driving amplitude, $\varepsilon=1.9\times10^{-3}$, which was below the threshold for these parameters, $\varepsilon_\text{cr}\approx2\times 10^{-3}$.
The agreement between the 3D (dotted magenta) and the approximated 1D (solid green) simulations is excellent.
We further study the threshold effect by scanning $\varepsilon$ for a fixed $\alpha$ and finding the critical value, $\varepsilon_\text{cr}$, that separates the captured and non-captured solutions.
The 3D simulations results (dots in the inset of Fig.~\ref{fig: capture_scatter_power_law}) well agree with the theoretical scaling, $\varepsilon_\text{cr}\propto\alpha^{3/4}$, which was developed for the 1D autoresonant pendulum \cite{fajans2001autoresonantT}, for three orders of magnitude (solid line).


A chirp cycle from $\omega_\text{d}\approx\omega_B$ to the lower limit for the loss cone crossing, $\omega_\text{d}=0$, lasts 
\begin{equation}
\Delta\, t=\frac{\Delta\omega_\text{d}}{\alpha\omega_0^2}\approx\frac{1}{\alpha \omega_0},
\end{equation}
where for the above parameters $\Delta\,t \approx 10$~ms.
This short time scale and the effect of automatic capture suggest using AR for removing  $\alpha$ particles from mirror (fusion) machines.
However, for efficient removal, a significant phase-space volume of initial conditions must be captured into AR, and the chirp duration must be shorter than a typical fusion burn time.
To estimate the phase space efficiency of the AR removal, we numerically solved the 3D dynamics of $10^4$ ($1$~MeV) $\alpha$ particles confined in the mirror.
The initial velocity directions were sampled from an isotropic distribution for a good phase space covering, while the system parameters were as in the example of Fig.~\ref{fig: capture_scatter_power_law}.
The simulation time was chosen such that the driving frequency, $\omega_\text{d}$, begins at $0.7~$MHz (above the linear resonance $\omega_\text{d}=\omega_B$), chirps down to $260$~KHz because, practically, autoresonant particles escape before $\omega_\text{d}=0$ due to nonlinearity and stochasticity near the separatrix \cite{Lazar_proc_2005}. 

\begin{figure}[t]
    \includegraphics[clip, trim=.0cm 1.6cm 0.0cm 0.95cm,
    width=1.0\linewidth]{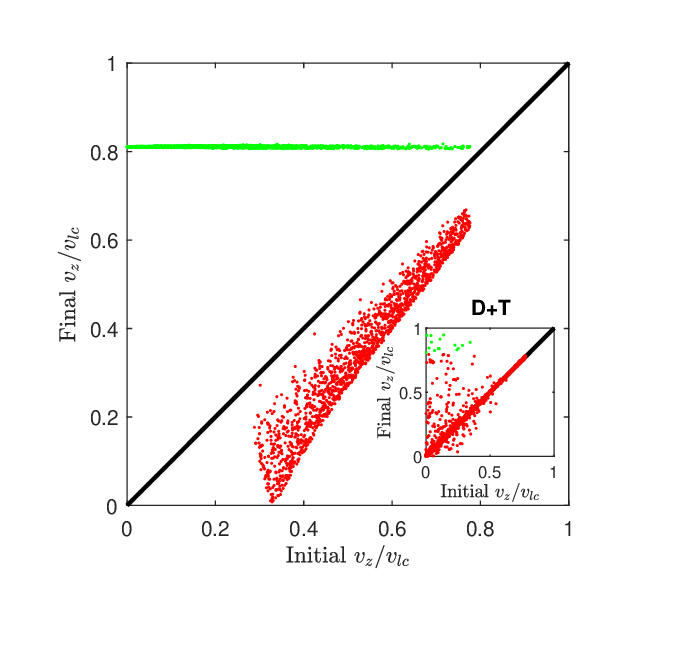}
    \caption{Final versus initial axial velocities normalized by the loss cone velocity, $v_\text{lc}$, of autoresonant $\alpha$ particles (green dots) and non-captured particles (red dots). 
    Inset: same plot for DT particles. 
    }
    \label{fig: MC_initial_final_vel_DT_1D}
\end{figure}

Fig.~\ref{fig: MC_initial_final_vel_DT_1D} presents the change (final vs. initial) in the axial velocity, $v_z$, normalized by the loss cone escape velocity, $v_\text{lc}=2\sqrt{\mu B_0/m}$, under the influence of the chirped drive for particles starting on the mid-plane, $z=0$, but with different $v_z$. 
In these calculations, a particle is said to be captured into AR if its axial velocity exceeds $0.8\,v_\text{lc}$, where the last mile outward can be supported by Coulomb collisions or nonresonant stochastic kicks induced by the successive drive cycles.
The figure demonstrates that all particles starting with small initial axial velocities, $v_\text{z} \le 0.3\, v_\text{lc}$, are captured into AR as expected in the automatic capture regime. 
In contrast, only part of the particles with higher initial $v_\text{z}$ are captured and can be associated with the stochastic capture regime \cite{Neishtadt_PRE_2005,Armon2016T}. 
Constructively, non-captured particles (red dots) are scattered towards lower axial velocities, making them more susceptible to AR removal for the successive chirping cycle.
At the same time, the DT fuel particles stay mostly unaffected by the chirped drive due to their lower mirror-bouncing frequencies as they are significantly less energetic ($\sim 15$~keV), so only a small fraction of their Maxwellian distribution have a sufficiently large $\mu$ for having $\omega_{B}$ in the AR chirping bandwidth.
This prediction is supported by the simulation results presented in the inset of Fig.~\ref{fig: MC_initial_final_vel_DT_1D}, showing that most of the DT fuel population remains close to the no-change line (black), $v_\text{final}=v_\text{initial}$.

\begin{figure}[t]
    \includegraphics[clip, trim=.7cm 0.cm 1.4cm 0.8cm,
    width=1\linewidth]{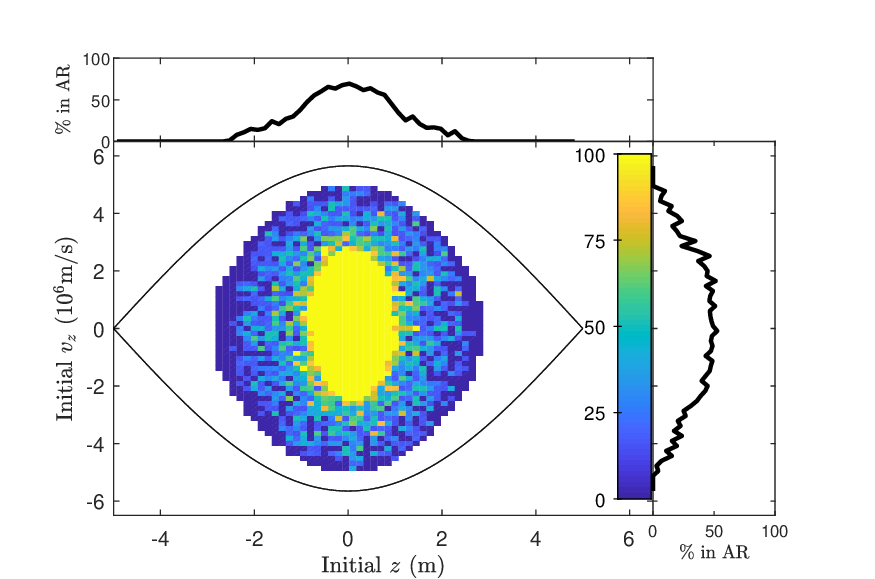}
    \caption{Autoresonant capture probability of on-axis $\alpha$ particles in the $z-v_z$ initial condition phase space. 
    The black lines denote the mirror separatrix.
    The outside panels display the marginal capture probability distribution in the $v_z$ (right) and $z$ (top) axes.}
    \label{fig: MC_z_initial_vel_15000_1D}
\end{figure}

 
We generalize the study from mid-plane initial conditions to particles starting at the $z-v_z$ phase space but still with a gyrocenter on the mirror axis. 
In Fig.~\ref{fig: MC_z_initial_vel_15000_1D}, $10^4$ particles were randomly sampled 
and binned, where, in each bin (pixel), the capture probability was calculated as the fraction of particles captured into AR in the bin. 
The simulation results demonstrate the existence of automatic (the central yellow elliptic region) and stochastic (the outer ring) capture into resonance regimes. 
It was also found that the automatic capture phase space volume increases with $\varepsilon$ (not presented).

\begin{figure}[tb]
    \includegraphics[clip, trim=0.85cm 0.1cm 1.4cm 1.0cm, width=1\linewidth]{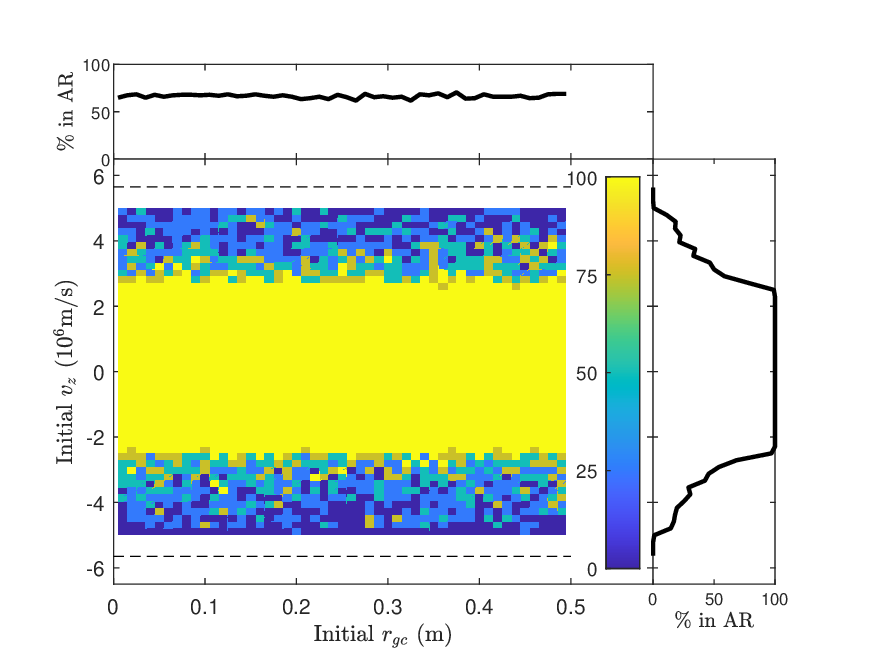}
    \caption{AR capture probability of off-axis $\alpha$ particles in the $r_{gc}-v_z$ initial condition phase space.
    The dashed black lines indicate the loss cone velocities. 
    The outside panels display the marginal capture probability distribution in the $v_z$ (right) and $r_{gc}$ (top) axes.
    }
    \label{fig: off_axis_particles}
\end{figure}

Finally, we study how the chirped drive affects $\alpha$ particles starting off-axis, i.e., gyrating around $r_{gc}>0$.
In the calculations presented in Fig. \ref{fig: off_axis_particles}, the sensitivity of the capture probability to the initial off-axis distance was tested up to $0.5~$m. 
The simulation results (upper panel) show a weak dependency of the capture probability on the initial $r_{gc}$, implying the robustness of the AR control for a reasonable range of radii.
It is noted that most of the $\alpha$ particles are born near the center (axially and radially) of a prospective fusion mirror-based machine, where the typical radial length for system parameters considered here is about $0.15~$m \cite{bagryansky2015overview,bagryansky2015threefold}.
Therefore, the efficiency of a single AR removal cycle, which can be calculated by integrating the capture probability function (Figs.~\ref{fig: MC_initial_final_vel_DT_1D}$-$\ref{fig: MC_z_initial_vel_15000_1D}) over the initial particle distribution in $v_z-z-r_{gc}$ phase space, is expected to be high.

The temporal efficiency depends on comparing the time required to accumulate an acceptable amount of $\alpha$ particles in the reactor and the AR removal time. 
We estimate the $\alpha$ accumulation time by $f/n_\text{DT}\langle\sigma v \rangle$, where $f$ is the desired product to fuel particle ratio, $n_\text{DT}$ is the fuel density, and $\langle\sigma v \rangle$ is averaged fusion reactivity.
For example, for $f=10^{-3}$, and plasma temperature and density of $15$~keV and $10^{14}$~cm$^{-3}$, respectively, one finds a production time of $\sim25$~ms, which is sufficiently longer than the few ms AR chirp cycle that removes most of the accumulated $\alpha$ particles.


Collisions affect the capture into AR by increasing the threshold value, $\varepsilon_\text{cr}$ and broadening its width \cite{barth2009autoresonantT}. 
These effects can be overcome for weak collisions by increasing the driving amplitude, $\varepsilon$. 
Since the typical cooling time of an $\alpha$ particle is hundreds of milliseconds while the chirping time is less than $10~$ms, the considered system is weakly collisional. 

The total energy of autoresonant particles in a magnetic mirror increases while their magnetic moment is preserved.
Therefore, the energy required to extract a particle from the trap is $\Delta U_{AR}=(R-1) U_\perp-U_\parallel$, where $U_{\perp}$ and $U_\parallel$ are the initial perpendicular and parallel kinetic energies, respectively, as evaluated at the mirror mid-plane and $R=B_{\text{max}}/B_{\text{min}}$ is the mirror ratio. 
For the example above, $\Delta U_{AR} \le 2$~MeV, which is about $10\%$ of the total fusion energy produced per reaction. 
Thus, for a typical fusion power plant concept, the AR removal scheme will require a few tens of MW. 
For example, the required current to generate a driving magnetic field of $B_d=25$~mT by a single-turn coil of $10$~cm in diameter located at the machine's throat is about $2$~kA.
To deliver a power of, say, $20$~MW to the $\alpha$ particles, the driving electric voltage should be about $10$~kV. 
Antennas with similar RF parameters (and with frequencies that are even higher than $1$~MHz) were successfully demonstrated in generating a field reverse configuration \cite{hoffman2002tcs}, and for ion cyclotron resonance heating in toroidal systems \cite{monakhov2024assessment,patel2018initial, lamalle2013status}.
It is noted, though, that most of the energy of the removed particles can be recovered via a direct energy conversion scheme \cite{Moir1973,barr1982experimental}, so the total power balance of the AR removal method is expected to be good. 

Notably, $\alpha$ particles of different energies, ranging from several times the fuel temperature and up to $3.5$~MeV, can be autoresonantly removed provided an appropriate chirping frequency range. 
For a given system, the frequency lower limit should be optimized to avoid resonant influence on the tail of the energy distribution of the fuel particles.
Importantly, targeting lower energy $\alpha$ particles is energetically favorable.
We also note that while an efficient and eventual removal of $\alpha$ particles is essential for the continuous operation of DT reactors, the prompt removal is of an absolute necessity for p-$^{11}$B reactors \cite{kolmes2022wave} because of their low reactivity and to avoid unwanted, secondary neutronic reactions.
Fortunately, the spectral separation between the $\alpha$ and the $^{11}$B fuel particles enables utilizing AR for quick and efficient $\alpha$ removal in such advanced reactors.

As to collective effects, the suggested AR scheme is a small-amplitude axisymmetric perturbation, so the induced instabilities are expected to be limited or even smaller than in unperturbed systems.
Moreover, it is noted that various methods to mitigate instabilities in mirror machines (\eg the flute instability) \cite{post1987magnetic, Bagryansky_2011, WHAM_2023} may also work in the presence of an AR perturbation. 
Still, the detailed analysis of MHD instabilities and other collective effects is left for a future study.

In conclusion, $\alpha$ particles born in fusion mirror machines can be efficiently removed by an autoresonant control of their axial motion.
Because of the spectral separation between the species, a significant portion of the fusion products can be extracted while the fuel particles remain nearly unaffected.
This novel scheme is based on a 1D guiding-center theory and supported by 3D numerical simulations. 
Phase space Monte Carlo analysis, including off-axis particles, verified the effectiveness and robustness of the scheme.
The AR selective expulsion method is also expected to be useful for various applications in mirror systems, including fusion product removal in aneutronic reactors, space propulsion, and heavy impurities removal \cite{kolmes2018strategies}.
It remains to study the application of AR in more complex open-field configurations, including the field reversal configuration and the tandem machine, as well as parametric autoresonant manipulation of closed 'banana' orbits in toroidal systems such as tokamaks and stellarators.

The authors thank Robert G. Littlejohn and Nathaniel J. Fisch for helpful conversations. 
This work was supported by the PAZY Foundation, Grant No. 2020-191.


\end{document}